\documentclass[12pt,english]{article}
\usepackage[T1]{fontenc}
\usepackage[latin9]{inputenc}
\usepackage{geometry}
\usepackage{babel}
\usepackage{url}
\usepackage{amsmath}
\usepackage{amssymb}
\usepackage{setspace}
\usepackage[authoryear]{natbib}
\PassOptionsToPackage{normalem}{ulem}
\usepackage{ulem}
\usepackage[unicode=true]
 {hyperref}

\makeatletter

\providecommand{\tabularnewline}{\\}

\usepackage{setspace}
\usepackage{pdflscape}
\usepackage{graphicx}
\usepackage{url}

\makeatother

\begin{document}
\begin{singlespacing}
\title{Dyadic Regression}

\maketitle
\medskip{}

\noindent \begin{center}
{\large{}Bryan S. Graham}\footnote{{\footnotesize{}Department of Economics, University of California
- Berkeley, 530 Evans Hall \#3380, Berkeley, CA 94720-3880 and National
Bureau of Economic Research, }{\footnotesize{}\uline{e-mail:}}{\footnotesize{}
\href{http://bgraham\%40econ.berkeley.edu}{bgraham@econ.berkeley.edu},
}{\footnotesize{}\uline{web:}}{\footnotesize{} \url{http://bryangraham.github.io/econometrics/}.
}\\
{\footnotesize{}This is a draft chapter for the edited volume }\emph{\footnotesize{}The
Econometric Analysis of Network Data}{\footnotesize{} being prepared
by Aureo de Paula and myself. It is based upon lecture notes prepared
for a series of short courses on the econometrics of networks. I thank
Michael Jansson for several useful discussions as well as participants
in short courses in Olso, Norway (September, 2017), Hejnice, Czechia
(February 2018), St. Gallen, Switzerland (October 2018), Annweiler,
Germany (July 2019) and Prague, Czechia (August 2019) for useful feedback.
Portions of this material were also presented at an invited session
of the 2018 LACEA/LAMES meetings in Guayaqil, Ecuador. All the usual
disclaimers apply. Financial support from NSF grant SES \#1851647
is gratefully acknowledged.}}
\par\end{center}

\noindent \begin{center}
\medskip{}
\textsc{\large{}\uline{Initial Draft}}\textsc{\large{}: September
2018, }\textsc{\large{}\uline{This Draft}}\textsc{\large{}: August
2019}{\large\par}
\par\end{center}

\begin{center}
{\Large{}Abstract}{\Large\par}
\par\end{center}

{\footnotesize{}\medskip{}
}{\footnotesize\par}

Dyadic data, where outcomes reflecting pairwise interaction among
sampled units are of primary interest, arise frequently in social
science research. Regression analyses with such data feature prominently
in many research literatures (e.g., gravity models of trade). The
dependence structure associated with dyadic data raises special estimation
and, especially, inference issues. This chapter reviews currently
available methods for (parametric) dyadic regression analysis and
presents guidelines for empirical researchers.{\footnotesize{}\medskip{}
}{\footnotesize\par}

\qquad{}\textbf{JEL Classification: C14, C55, C57}

{\footnotesize{}\medskip{}
}{\footnotesize\par}

\qquad{}\textbf{Keywords:} Networks, Dyadic Regression, U-Statistics

\end{singlespacing}

\thispagestyle{empty} \pagebreak{}

Let $Y_{ij}$ equal total exports from country $i$ to country $j$
as in \citet{Tinbergen_SWE62}; here $i$ and $j$ are two of $N$
independent random draws from a common population. Let $W_{i}$ be
a vector of country attributes and $R_{ij}=r\left(W_{i},W_{j}\right)$
a vector of constructed dyad-specific attributes; $R_{ij}$ typically
includes the logarithm of both exporter and importer gross domestic
product (GDP), the physical distance between $i$ and $j$, as well
as other variables (e.g., indicators for sharing a land border or
belonging to a common customs union). The analyst, seeking to relate
$Y_{ij}$ and $R_{ij}$, posits the relationship
\begin{equation}
Y_{ij}=\exp\left(R_{ij}'\theta_{0}\right)A_{i}B_{j}V_{ij},\label{eq: gravity_poisson}
\end{equation}
with $A_{i}$, $B_{i}$ and $V_{ij}$ mean one random variables and
$\left\{ \left(V_{ij},V_{ji}\right)\right\} _{1\leq i\leq N-1,j>i}$
independent of $\left\{ W_{i},A_{i},B_{i}\right\} _{i=1}^{N}$ and
independently and identically distributed across the $\tbinom{N}{2}=\frac{1}{2}N\left(N-1\right)$
dyads. Here the $\left\{ A_{i}\right\} _{i=1}^{N}$ and $\left\{ B_{i}\right\} _{i=1}^{N}$
sequences correspond, respectively, to (unobserved) exporter and importer
heterogeneity terms. These terms are sometimes referred to as ``multilaterial
resistance'' terms by empirical trade economists. For example, a
high $A_{i}$ might reflect an unmodeled export orientation of an
economy or an undervalued currency. Similarly, a high $B_{i}$ might
capture unmodeled tastes for consumption. \citet{Head_Mayer_HIE14}
survey the gravity model of trade, including its theoretical foundations.

Conditional on the exporter and importer effects we have
\[
\mathbb{E}\left[\left.Y_{ij}\right|W_{i},W_{j},A_{i},B_{j}\right]=\exp\left(R_{ij}'\theta_{0}\right)A_{i}B_{j}.
\]
If, additionally, $\mathbb{E}\left[\left.\left(A_{i},B_{i}\right)'\right|W_{i}\right]=\left(1,1\right)'$,
such that $W_{i}$ does not covary with the exporter and importer
``multilaterial resistance'' terms\footnote{If, for example, a subset of $\mathbb{W}$ is associated with membership
in the World Trade Organization (WTO), then reasoning about this condition
involves asking whether countries belonging to the WTO have a greater
latent propensity to export or import? In what follows I entirely
defer consideration of these questions and focus solely on the inferential
issues raised by the network structure.}, then unconditional on $A_{i}$ and $B_{j}$ we have the \emph{dyadic
regression} function
\begin{equation}
\mathbb{E}\left[\left.Y_{ij}\right|W_{i},W_{j}\right]=\exp\left(R_{ij}'\theta_{0}\right).\label{eq: poisson_reg}
\end{equation}
Interpret (\ref{eq: poisson_reg}) as follows: draw countries $i$
and $j$ independently at random and record their values of $W_{i}$
and $W_{j}$. Given this information set what is the mean square error
(MSE) minimizing predictor of $Y_{ij}$? Equation (\ref{eq: poisson_reg})
gives a parametric form for this prediction/regression function. This
chapter surveys methods of estimation of, and inference on, $\theta_{0}$.

\citet{SantosSilva_Tenreyro_RESTAT06} recommended estimating $\theta_{0}$
by maximizing a Poisson pseudo log-likelihood with a conditional mean
function given by (\ref{eq: poisson_reg}) \citep[cf.][]{Gourieroux_et_al_EM84}.
For inference they constructed standard errors using the sandwich
formula of \citet{Huber_BP67}; implicitly assuming that the $\left\{ Y_{ij}\right\} _{1\leq i,j\leq N,i\neq j}$
are conditionally independent of one another given $\mathbf{W}=\left(W_{1},\ldots,W_{N}\right)'$.
In practice this conditional independence assumption, although routinely
made in the empirical trade literature \citep[e.g.,][]{Rose_AER04,Baldwin_Taglioni_JEI07},
is very unlikely to hold. Exports from, say, Japan to Korea likely
covary with those from Japan to the United States. This follows because
$A_{i}$ -- the Japan exporter effect -- drives Japanese exports
to both Korea and the United States. It is also possible that exports
from Japan to Korea may covary with those from Korea to Thailand;
perhaps because $A_{j}$ and $B_{j}$ -- the Korean exporter and
importer effects -- covary (as would be true if there exist common
unobserved drivers of Korean exporting and importing behavior).\footnote{Researchers also sometimes ``cluster'' on dyads \citep[e.g.,][]{SantosSilva_Tenreyro_AR10};
this assumes that the elements of $\left\{ \left(Y_{ij},Y_{ji}\right)\right\} _{1\leq i\leq N-1,j>i}$
are conditionally independent given covariates. While this allows
for dependence between, say, exports from Japan to the United States
and from the United States to Japan, it does not allow for dependence
between, say, exports from Japan to the United States and from Japan
to Canada.}

Loosely following \citet{Fafchamp_Gubert_JDE07} I call the above
patterns of dependence ``dyadic dependence'' or ``dyadic clustering''.
Consider two pairs of dyads, say $\left\{ i_{1},i_{2}\right\} $ and
$\left\{ j_{1},j_{2}\right\} $, if these dyads share an agent in
common -- for example $i_{1}=j_{1}$ -- then $Y_{i_{1}i_{2}}$ and
$Y_{j_{1}j_{2}}$ will covary. Failing to account for dependence of
this type will, typically, result in standard errors which are too
small and consequently more Type I errors in inference than is desired
\citep[e.g.,][]{Cameron_Miller_WP14,Aronow_et_al_PA17}.

In this chapter I describe how to estimate and conduct inference on
$\theta_{0}$ in a way that appropriately accounts for dependence
across dyads sharing a unit in common. Section \ref{sec: Population-and-sampling}
outlines the population and sampling framework. Section \ref{sec: Composite-likelihood}
introduces a composite maximum likelihood estimator. Section \ref{sec: Limit-distribution}
develops the asymptotic properties of this estimator and discusses
variance estimation. Sections \ref{sec: Empirical-illustration} presents
a small empirical illustration.

Dyadic data, where outcomes reflecting pairwise interaction among
sampled units are of primary interest, arise frequently in social
science research. Such data play central roles in contemporary empirical
trade and international relations research (see, respectively, \citet{Tinbergen_SWE62}
and \citet{Oneal_Russett_WP99}). They also feature in work on international
financial flows \citep{Portes_Rey_JIE05}, development economics \citep{Fafchamp_Gubert_JDE07},
and anthropology \citep{Apicella_et_al_Nat12} among other fields.
Despite their prominence in empirical work, the properties of extant
methods of estimation and inference for dyadic regression models are
not fully understood. Only recently have researchers begun to formally
study these methods \citep[e.g.,][]{Aronow_et_al_PA17,Menzel_arXiv17,Tabord-Meehan_JBES18,Davezies_et_al_arXiv2019}.
Some of the results presented in this chapter are novel, others, while
having antecedents going back decades, are not widely known among
empirical researchers. Section \ref{sec: Further-reading} ends the
chapter with a discussion of further reading (including historically
important references).

\section{\label{sec: Population-and-sampling}Population and sampling framework}

Let $i\in\mathbb{N}$ index agents in some (infinite) population of
interest. In what follows I will refer to agents as, equivalently,
nodes, vertices, units and/or individuals. Let $W_{i}\in\mathbb{W}=\left\{ w_{1},\ldots,w_{L}\right\} $
be an observable attribute which partitions this population into $L=\left|\mathbb{W}\right|$
subpopulations or ``types''; $\mathbb{N}\left(w\right)=\left\{ i\thinspace:\thinspace W_{i}=w\right\} $
equals the index set associated with the subpopulation where $W_{i}=w$.
While $L$ may be very large, the size of each subpopulation is assumed
infinite. In practice $\mathbb{W}$ will typically enumerate different
combinations of distinct agent-specific attributes (e.g., $W_{i}=w_{1}$
may correspond to former British colonies in the tropics with per
capita GDP below \$3,000). Heuristically we can think of $\mathbb{W}$
as consisting of the support points of an multinomial approximation
to a (possibly continuous) underlying covariate space as in \citet{Chamberlain_JE87}.

The indexing of agents within subpopulations homogenous in $W_{i}$
is arbitrary; from the standpoint of the researcher all vertices of
the same type are exchangeable. Similar exchangeability assumptions
underlie most cross-sectional microeconometric procedures. For each
(ordered) pair of agents -- or \emph{directed dyad -- }there exists
an outcome of interest $Y_{ij}\in\mathbb{Y}\subseteq\mathbb{R}$.
The first subscript in $Y_{ij}$ indexes the directed dyads \emph{ego,}
or ``sending'' agent, while the second its \emph{alter,} or ``receiving''
agent. The \emph{adjacency matrix} $\left[Y_{ij}\right]_{i,j\in\mathrm{\mathbb{N}}}$
collects all such outcomes into an (infinite) random array. Within-type
exchangeability of agents implies a particular form of joint exchangeability
of the adjacency matrix.

To describe this exchangeability condition let $\sigma_{w}:\mathbb{N}\rightarrow\mathbb{N}$
be any permutation of indices satisfying the restriction
\begin{equation}
\left[W_{\sigma_{w}\left(i\right)}\right]_{i\in\mathbb{N}}=\left[W_{i}\right]_{i\in\mathbb{N}}.\label{eq: relatively_exchangeable_permutation}
\end{equation}
Condition (\ref{eq: relatively_exchangeable_permutation}) restricts
relabelings to occur among agents of the same type (i.e., \emph{within}
the index sets $\mathbb{N}\left(w\right)$, $w\in\mathbb{W}$). Following
\citet{Crane_Towsner_JSL18} a network is \emph{relatively exchangeable}
with respect to $W$ (or $W$-exchangeable) if, for all permutations
$\sigma_{w}$,

\begin{equation}
\left[Y_{\sigma_{w}\left(i\right)\sigma_{w}\left(j\right)}\right]_{i,j\in\mathbb{N}}\overset{D}{=}\left[Y_{ij}\right]_{i,j\in\mathbb{N}}\label{eq: relatively_exchangeability}
\end{equation}
where $\overset{D}{=}$ denotes equality of distribution.

If we regard $\left[Y_{ij}\right]_{i,j\in\mathrm{\mathbb{N}}}$ as
a (weighted) directed network and $W_{i}$ as vertex $i$'s ``color'',
then (\ref{eq: relatively_exchangeability}) is equivalent to the
statement that all colored graph isomorphisms are equally probable.
Since there is nothing in the researcher's information set which justifies
attaching different probabilities to graphs which are isomorphic (as
vertex colored graphs) any probability model for the adjacency matrix
should satisfy (\ref{eq: relatively_exchangeability}). If $W_{i}$
encodes all the vertex information observed by the analyst, then $W$-exchangeability
is a natural \emph{a priori} modeling restriction.

Condition (\ref{eq: relatively_exchangeability}) allows for the invocation
of very powerful \citet{deFinetti_AN1931} type representation results
for random arrays. These results provide an ``as if'' (nonparametric)
data generating process for the network adjacency matrix. This, in
turn, facilitates various probabilistic calculations (e.g., computing
expectations and variances) and gives (tractable) structure to the
dependence across the elements of $\left[Y_{ij}\right]_{i,j\in\mathrm{\mathbb{N}}}$.

Let $\alpha$, $\left\{ U_{i}\right\} _{i\geq1}$ and $\left\{ \left(V_{ij},V_{ji}\right)\right\} _{i\geq1,j>i}$
be i.i.d. random variables. We may normalize $\alpha$, $U_{ij}$
and $V_{ij}$ to be $\mathcal{U}\left[0,1\right]$ -- uniform on
the unit interval -- without loss of generality. We do allow for
within-dyad dependence across $V_{ij}$ and $V_{ji}$; the role such
dependence will become apparent below. Next consider the random array
$\text{\ensuremath{\left[Y_{ij}^{*}\right]}}_{i,j\in\mathbb{N}}$
generated according to the rule
\begin{equation}
Y_{ij}\overset{def}{\equiv}\tilde{h}\left(\alpha,W_{i},W_{j},U_{i},U_{j},V_{ij}\right).\label{eq: nonparametric_DGP}
\end{equation}
Data generating process (DGP) (\ref{eq: nonparametric_DGP}) has a
number of useful features. First, any pair of outcomes, $Y_{i_{1}i_{2}}$
and $Y_{j_{1}j_{2}}$, sharing at least one index in common are dependent.
This holds true even conditional on their types $W_{i_{1}},W_{i_{2}},W_{j_{1}}$
and $W_{j_{2}}$. Second, if $Y_{i_{1}i_{2}}$ and $Y_{j_{1}j_{2}}$
share exactly one index in common, say $i_{1}=j_{2}$, then they are
independent if $U_{i_{1}}=U_{j_{2}},U_{i_{2}}$and $U_{j_{1}}$ are
additionally conditioned on. Third, if they share both indices in
common, as in $i_{1}=j_{2}$ and $i_{2}=j_{1}$, then there may be
dependence even conditional on $U_{i_{1}}=U_{j_{2}}$ and $U_{i_{2}}=U_{j_{1}}$
due to the within-dyad dependence across $V_{i_{1}i_{2}}$ and $V_{i_{2}i_{1}}$.
These patterns of structured dependence and conditional independence
will be exploited below to derive the limit distribution of parametric
dyadic regression coefficient estimates. \citet{Shalizi_LN16} helpful
calls models like (\ref{eq: nonparametric_DGP}) conditionally independent
dyad (CID) models (see also \citet{Chandrasekhar_Book15}).

\citet{Crane_Towsner_JSL18}, extending \citet{Aldous_JMA81} and
\citet{Hoover_WP79}, show that, for any random array $\left[Y_{ij}\right]_{i,j\in\mathbb{N}}$
satisfying (\ref{eq: relatively_exchangeability}), there exists another
array $\text{\ensuremath{\left[Y_{ij}^{*}\right]}}_{i,j\in\mathbb{N}}$,
generated according to (\ref{eq: nonparametric_DGP}), such that
\begin{equation}
\left[Y_{ij}\right]_{i,j\in\mathbb{N}}\overset{D}{\equiv}\left[Y_{ij}^{*}\right]_{i,j\in\mathbb{N}}.\label{eq: Aldous_Hoover}
\end{equation}

Rule (\ref{eq: nonparametric_DGP}) can therefore be regarded as a
nonparametric data generating process for $\left[Y_{ij}\right]_{i,j\in\mathrm{\mathbb{N}}}$.
Equation (\ref{eq: Aldous_Hoover}) implies that we may proceed `as
if' our $W$-exchangeable network was generated according to (\ref{eq: nonparametric_DGP}).
In the spirit of \citet{Diaconis_Janson_RM08} and \citet{Bickel_Chen_PNAS09}
and others, call $\tilde{h}:\left[0,1\right]\times\mathbb{W}^{2}\times\left[0,1\right]^{3}\rightarrow\mathbb{R}$
a \emph{graphon}. Here $\alpha$ is an unidentiable mixing parameter,
analogous to the one appearing in de Finetti's \citeyearpar{deFinetti_AN1931}
classic representation result for exchangeable binary sequences. Since
I will focus on inference which is conditional on the empirical distribution
of the data, $\alpha$ can be safely ignored and I will write $h\left(W_{i},W_{j},U_{i},U_{j},V_{ij}\right)\overset{def}{\equiv}\tilde{h}\left(\alpha,W_{i},W_{j},U_{i},U_{j},V_{ij}\right)$
in what follows \citep[cf.,][]{Bickel_Chen_PNAS09,Menzel_arXiv17}.

The \citet{Crane_Towsner_JSL18} representation result implies that
a very particular type of dependence structure is associated with
$W$-exchangeability. Namely, as discussed earlier, $Y_{i_{1}i_{2}}$
and $Y_{j_{1}j_{2}}$ are (conditionally) independent when $\left\{ i_{1},i_{2}\right\} $
and $\left\{ j_{1},j_{2}\right\} $ share no indices in common and
dependent when they do. This type of dependence structure, which is
very much analogous to that which arises in the theory U-Statistics,
is tractable and allows for the formulation of Laws of Large Numbers
and Central Limit Theorems. The next few sections will show how to
use this insight to develop asymptotic distribution theory for dyadic
regression.

\subsection*{Sampling assumption}

I will regard $\left[Y_{ij}\right]_{i,j\in\mathrm{\mathbb{N}}}$ as
an infinite random (weighted) graph, $G_{\infty},$ with nodes $\mathbb{N}$
and (weighted) edges given by the non-zero elements of $\left[Y_{ij}\right]_{i,j\in\mathrm{\mathbb{N}}}$.
Let $\mathcal{V}=\left\{ 1,\ldots,N\right\} $ be a random sample
of size $N$ from $\mathbb{N}$. Let $G_{N}=G_{\infty}\left[\mathit{\mathcal{V}}\right]$
be the subgraph indexed by $\mathcal{V}$. We assume that the observed
network corresponds to the one induced by a random sample of agents
from the larger (infinite) graph. The sampling distribution of any
statistic of $G_{N}$ is induced by this (perhaps hypothetical) random
sampling of agents from $G_{\infty}$.

If $G_{\infty}$ is relatively exchangeable, then $G_{N}$ will we
be as well. We can thus proceed `as if'
\[
Y_{ij}=h\left(W_{i},W_{j},U_{i},U_{j},V_{ij}\right)
\]
for $1\leq i,j\leq N$. In what follows we assume that we observe
$W_{i}$ for each sampled agent, and for each pair of sampled agents,
we observe both $Y_{ij}$ and $Y_{ji}$. The presentation here rules
out self loops (i.e.,$Y_{ii}\equiv0$), however incorporating them
is natural in some empirical settings and what follows can be adapted
to handle them. Similarly the extension to undirected outcomes, where
$Y_{ij}=Y_{ji}$, is straightforward.

\section{\label{sec: Composite-likelihood}Composite likelihood}

Let $f_{\left.Y_{12}\right|W_{1},W_{2}}\left(\left.Y_{12}\right|W_{1},W_{2};\theta\right)$
be a parametric family for the conditional density of $Y_{12}$ given
$W_{1}$ and $W_{2}$. This family is chosen by the researcher. Let
$l_{12}\left(\theta\right)$ denote the corresponding log-likelihood.
As an example to help fix ideas, return to the variant of the gravity
model of trade introduced in the introduction. Following \citet{SantosSilva_Tenreyro_RESTAT06}
we set
\[
l_{12}\left(\theta\right)=Y_{12}R_{12}'\theta-\exp\left(R_{12}'\theta\right),
\]
which equals (up to a term not varying with $\theta$) the log likelihood
of a Poisson random variable $Y_{12}$ with mean $\exp\left(R_{12}'\theta\right)$,
and choose $\hat{\theta}$ to maximize
\begin{equation}
L_{N}\left(\theta\right)=\frac{1}{N}\frac{1}{N-1}\sum_{i}\sum_{j\neq i}l_{ij}\left(\theta\right).\label{eq: composite_log_likelihood}
\end{equation}
The maximizer of (\ref{eq: composite_log_likelihood}) coincides with
a maximum likelihood estimate based upon the assumption that $\left[Y_{ij}\right]_{1\leq i,j\leq N,i\neq j}$
are independent Poisson random variables conditional on $\mathbf{W}=\left(W_{1},\ldots,W_{N}\right)'.$

In practice, trade flows are unlikely to be well-described by a Poisson
distribution and independence of the summands in (\ref{eq: composite_log_likelihood})
is even less likely. As discussed earlier any two summands in (\ref{eq: composite_log_likelihood})
will be dependent if they share an index in common. The likelihood
contribution associated with exports from Vanuatu to Fiji is not independent
of that associated with exports from Fiji to Bangladesh. Dependencies
of this type mean that proceeding `as if' (\ref{eq: composite_log_likelihood})
is a correctly specified log-likelihood (or even an M-estimation criterion
function) will lead to incorrect inference.

If there exists some $\theta_{0}$ such that $f_{\left.Y_{12}\right|W_{1},W_{2}}\left(\left.Y_{12}\right|W_{1},W_{2};\theta_{0}\right)$
is the true density, then (\ref{eq: nonparametric_DGP}) corresponds
to what is called a \emph{composite} likelihood \citep[e.g.,][]{Lindsay_CM88,Cox_Reid_BM04,Belio_Varin_SM05}.
Because it does not correctly reflect the dependence structure across
dyads, (\ref{eq: nonparametric_DGP}) is not a correctly specified
log-likelihood function in the usual sense. If, however, the marginal
density of $\left.Y_{ij}\right|W_{i},W_{j}$ is correctly specified,
then $\hat{\theta}$ will generally be consistent for $\theta_{0}$.
That is we may have that
\[
f_{\left.Y_{12}\right|W_{1},W_{2}}\left(\left.Y_{12}\right|W_{1},W_{2}\right)=f_{\left.Y_{12}\right|W_{1},W_{2}}\left(\left.Y_{12}\right|W_{1},W_{2};\theta_{0}\right)
\]
for some $\theta_{0}\in\Theta$ (i.e., the marginal likelihood is
correctly specified), but it \emph{is not} the case that, setting
$\mathbf{Y}=\left[Y_{ij}\right]_{1\leq i,j\leq N,i\neq j}$,
\[
f_{\left.\mathbf{Y}\right|\mathbf{W}}\left(\left.\mathbf{Y}\right|\mathbf{W}\right)=\prod_{1\leq i,j\leq N,i\neq j}f_{\left.Y_{12}\right|W_{1},W_{2}}\left(\left.Y_{ij}\right|W_{i},W_{j};\theta_{0}\right),
\]
due to dependence across dyads sharing agents in common (i.e., the
joint likelihood is not correctly specified). A composite log-likelihood
is constructed by summing together a collection of component log-likelihoods;
each such component is a log-likelihood for a portion of the sample
(in this case a single \emph{directed} dyad) but, because the joint
dependence structure may not be modeled appropriately, the summation
of all these components may not be the correct log likelihood for
the sample as a whole.

If the marginal likelihood is itself misspecified, then (\ref{eq: nonparametric_DGP})
corresponds to what might be called a pseudo-composite-log-likelihood;
``pseudo'' in the sense of \citet{Gourieroux_et_al_EM84} and ``composite''
in the sense of \citet{Lindsay_CM88}. In what follows I outline how
to conduct inference on the probability limit of $\hat{\theta}$ (denoted
by $\theta_{0}$ in all cases); the interpretation of this limit will,
of course, depend on whether the pairwise likelihood is misspecified
or not. In the context of the \citet{SantosSilva_Tenreyro_RESTAT06}
gravity model example, if the true conditional mean equals $\exp\left(R_{ij}'\theta_{0}\right)$
for some $\theta_{0}\in\Theta$, then $\hat{\theta}$ will be consistent
for it (under regularity conditions). The key challenge is to characterize
this estimate's sampling precision.

\section{\label{sec: Limit-distribution}Limit distribution}

To characterize the limit properties of $\hat{\theta}$ begin with
a mean value expansion of the first order condition associated with
the maximizer of (\ref{eq: composite_log_likelihood}). This yields,
after some re-arrangement,
\[
\sqrt{N}\left(\hat{\theta}-\theta_{0}\right)=\left[-H_{N}\left(\bar{\theta}\right)\right]^{+}\sqrt{N}S_{N}\left(\theta_{0}\right)
\]
with $\bar{\theta}$ a mean value between $\hat{\theta}_{\mathrm{}}$
and $\theta_{0}$ which may vary from row to row, the $+$ superscript
denoting a Moore-Penrose inverse, and a ``score'' vector of
\begin{equation}
S_{N}\left(\theta\right)=\frac{1}{N}\frac{1}{N-1}\sum_{i}\sum_{j\neq i}s_{ij}\left(Z_{ij},\theta\right)\label{eq: S_N}
\end{equation}
with $s\left(Z_{ij},\theta\right)=\partial l_{ij}\left(\theta\right)/\partial\theta$
for $Z_{ij}=\left(Y_{ij},W_{i}',W_{j}'\right)'$ and $H_{N}\left(\theta\right)=\frac{1}{N}\frac{1}{N-1}\sum_{i}\sum_{j\neq i}\frac{\partial^{2}l_{ij}\left(\theta\right)}{\partial\theta\partial\theta'}$.
In what follows I will just assume that $H_{N}\left(\bar{\theta}\right)\overset{p}{\rightarrow}\Gamma_{0}$,
with $\Gamma_{0}$ invertible (see \citet{Graham_EM17} for a formal
argument in a related setting and \citet{Eagleson_Weber_MP78} and
\citet{Davezies_et_al_arXiv2019} for more general results).

If the Hessian matrix converges in probability to $\Gamma_{0}$, as
assumed, then
\[
\sqrt{N}\left(\hat{\theta}-\theta_{0}\right)=\Gamma_{0}^{-1}\sqrt{N}S_{N}\left(\theta_{0}\right)+o_{p}\left(1\right)
\]
so that the asymptotic sampling properties of $\sqrt{N}\left(\hat{\theta}_{\mathrm{}}-\theta_{0}\right)$
will be driven by the behavior of $\sqrt{N}S_{N}\left(\theta_{0}\right)$.
As pointed out by \citet{Fafchamp_Gubert_JDE07} and others, (\ref{eq: S_N})
is not a sum of independent random variables, hence a basic central
limit theorem (CLT) cannot be (directly) applied.

My analysis of $\sqrt{N}S_{N}\left(\theta_{0}\right)$ borrows from
the theory of U-Statistics \citep[e.g.,][]{Ferguson_UStat05,vanderVaart_ASBook00}.
To make these connections clear it is convenient to re-write $S_{N}\left(\theta_{0}\right)$
as
\[
S_{N}\left(\theta\right)=\binom{N}{2}^{-1}\sum_{i<j}\left\{ \frac{s\left(Z_{ij},\theta\right)+s\left(Z_{ji},\theta\right)}{2}\right\} 
\]
where $\sum_{i<j}\overset{def}{\equiv}\sum_{i=1}^{N-1}\sum_{j=i+1}^{N}$.

Let $s_{ij}\overset{def}{\equiv}s\left(Z_{ij},\theta_{0}\right)$,
$S_{N}=S_{N}\left(\theta_{0}\right)$ and $\bar{s}\left(w,u,w',u'\right)=\mathbb{E}\left[\left.s_{12}\right|W_{1}=w,U_{1}=u,W_{2}=w',U_{2}=u'\right];$
next decompose $S_{N}$ as follows
\[
S_{N}=U_{N}+V_{N},
\]
where $U_{N}$ equals the projection of $S_{N}$ onto $\mathbf{W}=\left[W_{i}\right]_{1\leq i\leq N}$
\emph{and} $\mathbf{U}=\left[U_{i}\right]_{1\leq i\leq N}$:
\begin{equation}
U_{N}=\mathbb{E}\left[\left.S_{N}\right|\mathbf{W},\mathbf{U}\right]=\binom{N}{2}^{-1}\sum_{i<j}\frac{\bar{s}\left(W_{i},U_{i},W_{j},U_{j}\right)+\bar{s}\left(W_{j},U_{j},W_{i},U_{i}\right)}{2}\label{eq: U_N}
\end{equation}
and $V_{N}=S_{N}-U_{N}$ is the corresponding projection error:
\begin{equation}
V_{N}=\binom{N}{2}^{-1}\sum_{i<j}\frac{\left[s\left(Z_{ij},\theta\right)-\bar{s}\left(W_{i},U_{i},W_{j},U_{j}\right)\right]+\left[s\left(Z_{ji},\theta\right)-\bar{s}\left(W_{j},U_{j},W_{i},U_{i}\right)\right]}{2}.\label{eq: V_N}
\end{equation}
Observe that $U_{N}$ and $V_{N}$ are uncorrelated by construction.
Furthermore $U_{N}$ is a U-statistic, albeit defined -- partially
-- in terms of the latent variable $U_{i}$. Although we can not
numerically evaluate $U_{N}$, we can characterize is sampling properties
as $N\rightarrow\infty$. In order to do so we further decompose $U_{N}$
into a Hájek projection and a second remainder term:

\[
U_{N}=U_{1N}+U_{2N}
\]
where, defining $\bar{s}_{1}^{e}\left(w,u\right)=\mathbb{E}\left[\bar{s}\left(w,u,W_{1},U_{1}\right)\right]$
and $\bar{s}_{1}^{a}\left(w,u\right)=\mathbb{E}\left[\bar{s}\left(W_{1},U_{1},w,u\right)\right]$,
\begin{align*}
U_{1N}= & \frac{2}{N}\sum_{i=1}^{N}\frac{\bar{s}_{1}^{e}\left(W_{i},U_{i}\right)+\bar{s}_{1}^{a}\left(W_{i},U_{i}\right)}{2}\\
U_{2N}= & \binom{N}{2}^{-1}\sum_{i<j}\left\{ \frac{\bar{s}\left(W_{i},U_{i},W_{j},U_{j}\right)+\bar{s}\left(W_{j},U_{j},W_{i},U_{i}\right)}{2}\right.\\
 & \left.-\frac{\bar{s}_{1}^{e}\left(W_{i},U_{i}\right)+\bar{s}_{1}^{a}\left(W_{i},U_{i}\right)}{2}-\frac{\bar{s}_{1}^{e}\left(W_{j},U_{j}\right)+\bar{s}_{1}^{a}\left(W_{j},U_{j}\right)}{2}\right\} 
\end{align*}
The superscript in $\bar{s}_{1}^{e}\left(W_{i},U_{i}\right)$ stands
for `ego' since $\bar{s}_{1}^{e}\left(W_{1},U_{1}\right)=\mathbb{E}\left[\left.\bar{s}\left(W_{1},U_{1},W_{2},U_{2}\right)\right|W_{1},U_{1}\right]$
corresponds to the expected value of a (generic) dyad's contribution
to the composite likelihood's score vector holding its ego's attributes
fixed. Similarly the superscript in $\bar{s}_{1}^{a}\left(W_{i},U_{i}\right)$
stands for `alter', since it is her attributes being held fixed in
that average.

Putting things together yields the score decomposition
\[
S_{N}=\overset{\text{(First) Projection onto \textbf{W} and \textbf{U}}}{\overbrace{\underset{\text{(Second) Hájek Projection}}{\underbrace{U_{1N}}}+\underset{\text{(Second) Projection Error}}{\underbrace{U_{2N}}}}}+\overset{\text{(First) Projection Error}}{\overbrace{V_{N}}}.
\]
The limit distribution of $\sqrt{N}\left(\hat{\theta}-\theta_{0}\right)$
depends on the joint behavior of $U_{1N}$, $U_{2N}$ and $V_{N}$
as $N\rightarrow\infty$. A similar type of double projection argument
was utilized by \citet{Graham_EM17} to characterize the limit distribution
of the Tetrad Logit estimator.\footnote{It is also implicit in the analysis of \citet{Bickel_et_al_AS11}.}
The analyses of \citet{Menzel_arXiv17} and \citet{Graham_Niu_Powell_WP2019}
both utilize a similar decomposition.

\subsection*{Variance calculation}

In this section I first derive the sampling variance of $\sqrt{N}\left(\hat{\theta}-\theta_{0}\right)$
and then provide an interpretation of it. I begin by calculating the
variance of $S_{N}:$
\[
\mathbb{V}\left(S_{N}\right)=\mathbb{V}\left(U_{1N}\right)+\mathbb{V}\left(U_{2N}\right)+\mathbb{V}\left(V_{N}\right).
\]

Let
\[
\Sigma_{q}=\mathbb{C}\left(\bar{s}\left(W_{i_{1}},U_{i_{1}},W_{i_{2}},U_{i_{2}}\right)+\bar{s}\left(W_{i_{2}},U_{i_{2}},W_{i_{1}},U_{i_{1}}\right),\bar{s}\left(W_{j_{1}},U_{j_{1}},W_{j_{2}},U_{j_{2}}\right)+\bar{s}\left(W_{j_{2}},U_{j_{2}},W_{j_{1}},U_{j_{1}}\right)\right)
\]
when the dyads $\left\{ i_{1},i_{2}\right\} $ and $\left\{ j_{1},j_{2}\right\} $
share $q=0,1,2$ indices in common. A \citet{Hoeffding_AMS48} variance
decomposition gives
\begin{align*}
\mathbb{V}\left(U_{N}\right) & =\mathbb{V}\left(U_{1N}\right)+\mathbb{V}\left(U_{2N}\right)\\
 & \frac{4}{N}\Sigma_{1}+\frac{2}{N\left(N-1\right)}\left(\Sigma_{2}-\Sigma_{1}\right).
\end{align*}
Direct calculation yields (see Appendix \ref{app: Derivations})
\begin{align}
\Sigma_{1} & \overset{def}{\equiv}\mathbb{V}\left(\frac{\bar{s}_{1}^{e}\left(W_{1},U_{1}\right)+\bar{s}_{1}^{a}\left(W_{1},U_{1}\right)}{2}\right)\label{eq: SIGMA1}\\
 & =\frac{\Omega_{12,13}+2\Omega_{12,31}+\Omega_{21,31}}{4}\nonumber 
\end{align}
with
\[
\Omega_{i_{1}i_{2},j_{1}j_{2}}=C\left(\bar{s}\left(W_{i_{1}},U_{i_{1}},W_{i_{2}},U_{i_{2}}\right),\bar{s}\left(W_{j_{1}},U_{j_{1}},W_{j_{2}},U_{j_{2}}\right)\right).
\]

Similarly we have
\begin{align}
\Sigma_{2} & =\mathbb{V}\left(\frac{\bar{s}\left(W_{1},U_{1},W_{2},U_{2}\right)+\bar{s}\left(W_{2},U_{2},W_{1},U_{1}\right)}{2}\right)\label{eq: SIGMA2}\\
 & =\frac{\Omega_{12,12}+\Omega_{12,21}}{2}\nonumber 
\end{align}
and, in an abuse of notation, letting $\Sigma_{3}\overset{def}{\equiv}\mathbb{V}\left(\sqrt{\tbinom{N}{2}}V_{N}\right),$
\begin{align}
\Sigma_{3} & =\mathbb{E}\left[\frac{\Delta_{12,12}\left(W_{1},U_{1},W_{2},U_{2}\right)+\Delta_{12,21}\left(W_{1},U_{1},W_{2},U_{2}\right)}{2}\right]\label{eq: SIGMA3}\\
 & =\frac{\bar{\Delta}_{12,12}+\bar{\Delta}_{12,21}}{2}\nonumber 
\end{align}
where
\begin{align*}
\Delta_{12,12}\left(W_{1},U_{1},W_{2},U_{2}\right) & =\mathbb{V}\left(\left.s\left(Z_{12},\theta\right)\right|W_{1},U_{1},W_{2},U_{2}\right)\\
\Delta_{12,21}\left(W_{1},U_{1},W_{2},U_{2}\right) & =\mathbb{E}\left[\left.s\left(Z_{12},\theta\right)s\left(Z_{21},\theta\right)'\right|W_{1},U_{1},W_{2},U_{2}\right].
\end{align*}

From (\ref{eq: SIGMA1}), (\ref{eq: SIGMA2}) and (\ref{eq: SIGMA3})
we have, collecting terms, a variance of $S_{N}$ equal to
\begin{align}
\mathbb{V}\left(S_{N}\right)= & \mathbb{V}\left(U_{1N}\right)+\mathbb{V}\left(U_{2N}\right)+\mathbb{V}\left(V_{N}\right)\label{eq: Variance_of_S_N}\\
 & \frac{4}{N}\Sigma_{1}+\frac{2}{N\left(N-1\right)}\left(\Sigma_{2}-2\Sigma_{1}\right)+\frac{2}{N\left(N-1\right)}\Sigma_{3}\nonumber \\
= & \left(\Omega_{12,13}+2\Omega_{12,31}+\Omega_{21,31}\right)\left(\frac{N-2}{N-1}\right)\nonumber \\
 & +\frac{1}{N-1}\left(\Omega_{12,12}+\bar{\Delta}_{12,12}+\Omega_{12,21}+\bar{\Delta}_{12,21}\right).\nonumber 
\end{align}
To understand (\ref{eq: Variance_of_S_N}) note that there are exactly
$\dbinom{N}{2}\dbinom{2}{1}\dbinom{N-2}{1}=N\left(N-1\right)\left(N-2\right)$
pairs of dyads sharing one agent in common. Consequently, applying
the variance operator to $S_{N}$ yields a total of $N\left(N-1\right)\left(N-2\right)$
non-zero covariance terms across the $\dbinom{N}{2}$ summands in
$S_{N}$. It is these covariance terms which account for the leading
term in (\ref{eq: Variance_of_S_N}). The second and third terms in
(\ref{eq: Variance_of_S_N}) arise from the $\dbinom{N}{2}$ variances
of the summands in $S_{N}$. Indeed, it is helpful to note that
\begin{align*}
\Sigma_{2} & =\mathbb{V}\left(\mathbb{E}\left[\left.\frac{s\left(Z_{12},\theta\right)+s\left(Z_{21},\theta\right)}{2}\right|W_{1},U_{1},W_{2},U_{2}\right]\right)\\
\Sigma_{3} & =\mathbb{E}\left[\mathbb{V}\left(\left.\frac{s\left(Z_{12},\theta\right)+s\left(Z_{21},\theta\right)}{2}\right|W_{1},U_{1},W_{2},U_{2}\right)\right]
\end{align*}
and hence that
\begin{equation}
\mathbb{V}\left(\frac{s\left(Z_{12},\theta\right)+s\left(Z_{21},\theta\right)}{2}\right)=\Sigma_{2}+\Sigma_{3}.\label{eq: SIGMA2_SIGMA3}
\end{equation}

Although it may be that $\Sigma_{2}+\Sigma_{3}\geq\Sigma_{1}$ (in
a positive definite sense), the larger number of non-zero covariance
terms generated by applying the variance operator to $S_{N}$ contributes
more to its variability, than the smaller number of own variance terms.
Inspecting (\ref{eq: Variance_of_S_N}) it is clear that the multiplying
by $\sqrt{N}$ stabilizes the variance such that
\[
\mathbb{V}\left(\sqrt{N}S_{N}\right)=4\Sigma_{1}+O\left(N^{-1}\right)
\]
and hence
\[
\mathbb{V}\left(\sqrt{N}\left(\hat{\theta}-\theta\right)\right)\rightarrow4\left(\Gamma'\Sigma_{1}^{-1}\Gamma\right)^{-1}
\]
as $N\rightarrow\infty$.

If a researcher uses standard software, for example a Poisson regression
program, to maximize the composite log-likelihood (\ref{eq: composite_log_likelihood})
and then chooses to report robust \citet{Huber_BP67} type standard
errors, this corresponds to assuming that
\[
\Omega_{12,13}=\Omega_{12,31}=\Omega_{21,31}=\Omega_{12,21}=\bar{\Delta}_{12,21}=0.
\]
This approach would ignore the dominant variance term and part of
the higher order term as well. If, instead, the researcher clustered
her standard errors on dyads, as in, for example, \citet{SantosSilva_Tenreyro_AR10},
then this corresponds to assuming that
\[
\Omega_{12,13}=\Omega_{12,31}=\Omega_{21,31}=0
\]
but allowing $\Omega_{12,21}$ and/or $\bar{\Delta}_{12,21}$ to differ
from zero. This approach would still erroneously ignore the dominant
variance term. In both cases reported confidence intervals are likely
to undercover the true parameter; perhaps by a substantial margin.
This is shown, by example, via Monte Carlo simulation below.

\subsection*{Variance estimation}

\citet{Graham_HBE18} provides a comprehensive discussion of variance
estimation for dyadic regression. One approach to variance estimation
he reviews shows that $\Sigma_{1}$ can be estimated by the analog
covariance estimate
\begin{align*}
\hat{\Sigma}_{1}= & \frac{1}{4}\frac{2}{N\left(N-1\right)\left(N-1\right)}\sum_{i=1}^{N-2}\sum_{j=i+1}^{N-1}\sum_{k=j+1}^{N}\left\{ \left(\hat{s}_{ij}+\hat{s}_{ji}\right)\left(\hat{s}_{ik}+\hat{s}_{ki}\right)'\right.\\
 & \left.\left(\hat{s}_{ij}+\hat{s}_{ji}\right)\left(\hat{s}_{jk}+\hat{s}_{kj}\right)'+\left(\hat{s}_{ik}+\hat{s}_{ki}\right)\left(\hat{s}_{jk}+\hat{s}_{kj}\right)'\right\} ,
\end{align*}
where the summation is over all triads in the sampled network. Each
triad can itself be partitioned into three different pairs of dyads,
each sharing an agent in common.

It turns out, as inspection of (\ref{eq: SIGMA2_SIGMA3}) suggests,
it is easiest to estimate the sum of $\Sigma_{2}$ and $\Sigma_{3}$
jointly by
\begin{align*}
\widehat{\Sigma_{2}+\Sigma_{3}} & =\frac{1}{4}\frac{2}{N\left(N-1\right)}\sum_{i=1}^{N-1}\sum_{j=i+1}^{N}\left(\hat{s}_{ij}+\hat{s}_{ji}\right)\left(\hat{s}_{ij}+\hat{s}_{ji}\right)'.
\end{align*}
 The Jacobian matrix, $\Gamma_{0}$, may be estimated by $-H_{N}\left(\hat{\theta}\right)$,
which is typically available as a by-product of estimation in most
commercial software. Putting things together gives a variance estimate
of
\begin{equation}
\mathbb{\hat{V}}\left(\sqrt{N}\left(\hat{\theta}-\theta_{0}\right)\right)=\hat{\Gamma}^{-1}\left(4\hat{\Sigma}_{1}+\frac{2}{N-1}\left(\widehat{\Sigma_{2}+\Sigma_{3}}-2\hat{\Sigma}_{1}\right)\right)\left(\hat{\Gamma}^{-1}\right)'.\label{eq: FG_Variance_Estimate}
\end{equation}
\citet{Graham_HBE18} shows that (\ref{eq: FG_Variance_Estimate})
is numerically equivalent, up to a finite sample correction, to the
variance estimator proposed by \citet{Fafchamp_Gubert_JDE07}. This
variance estimator includes estimates of asymptotically negligible
terms. Although these terms are negligible when the sample is large
enough, in practice they may be sizable in real world settings.

\subsection*{Limit distribution}

The variance calculations outlined above imply that $\sqrt{N}S_{N}=\sqrt{N}U_{1N}+o_{p}\left(1\right)$
and hence that

\[
\sqrt{N}\left(\hat{\theta}-\theta_{0}\right)=\Gamma_{0}^{-1}\sqrt{N}U_{1N}+o_{p}\left(1\right).
\]
Since $U_{1N}$ is the sum of i.i.d. random variables a CLT gives
\begin{equation}
\sqrt{N}\left(\hat{\theta}_{\mathrm{}}-\theta_{0}\right)\overset{D}{\rightarrow}\mathcal{N}\left(0,4\left(\Gamma_{0}'\Sigma_{1}^{-1}\Gamma_{0}\right)^{-1}\right),\label{eq: theta_hat_dr_limit_distribution}
\end{equation}
The variance expression, equation (\ref{eq: Variance_of_S_N}), indicates
that inference based upon the limit distribution (\ref{eq: theta_hat_dr_limit_distribution})
would ignore higher order variance terms included in (\ref{eq: FG_Variance_Estimate}).
In practice, as has been shown in other contexts, an approach to inference
which incorporates estimates of these higher order variance terms
may result in inference with better size properties \citep[e.g.,][]{Graham_Imbens_Ridder_QE14,Cattaneo_et_al_ET14,Graham_Niu_Powell_WP2019}.
In practice I suggest using the normal reference distribution, but
with a variance estimated by (\ref{eq: FG_Variance_Estimate}), which
includes asymptotically negligible terms which may nevertheless be
large in real world samples.

\section{\label{sec: Empirical-illustration}Empirical illustration}

This section provides an example of a dyadic regression analysis using
the dataset constructed by João Santos Silva and Silvana Tenreyro
\citeyearpar{SantosSilva_Tenreyro_RESTAT06} in their widely-cited
paper ``The Log of Gravity''. This dataset, which as of the Fall
of 2019 was available for download at \url{http://personal.lse.ac.uk/tenreyro/LGW.html},
includes information on $N=136$ countries, corresponding to 18,360
directed trading relationships. Here I present a simple specification
which includes only the log of exporter and importer GDP, respectively
$\mathtt{lyex}$ and $\mathtt{lyim}$, as well as the log distance
($\mathtt{ldist}$) between the two trading countries. Maximizing
(\ref{eq: composite_log_likelihood}) yields a fitted regression function
of
\[
\mathbb{\hat{E}}\left[\left.Y_{ij}\right|W_{i},W_{j}\right]=\exp\left(\begin{array}{c}
-5.688\\
\left(1.9382\right)
\end{array}+\begin{array}{c}
0.9047\\
\left(0.0750\right)
\end{array}\begin{array}{c}
\mathcal{\mathtt{lyex}}\\
\\
\end{array}+\begin{array}{c}
0.8941\\
\left(0.0668\right)
\end{array}\begin{array}{c}
\mathtt{lyim}\\
\\
\end{array}+\begin{array}{c}
-0.5676\\
\left(0.0982\right)
\end{array}\begin{array}{c}
\mathtt{ldist}\\
\\
\end{array}\right).
\]
Standard errors which cluster on dyads, but ignore dependence across
dyads sharing a single agent in common, are reported in parentheses
below the coefficient estimates. Specifically these standard errors
coincide with square roots of the diagonal elements of
\begin{equation}
\frac{2}{N\left(N-1\right)}\hat{\Gamma}^{-1}\left(\widehat{\Sigma_{2}+\Sigma_{3}}\right)\left(\hat{\Gamma}^{-1}\right)'.\label{eq: iid_standard_errors}
\end{equation}
The coefficient estimates and reported standard errors are unremarkable
in the context of the empirical trade literature. I refer the reader
to \citet{SantosSilva_Tenreyro_RESTAT06} or \citet{Head_Mayer_HIE14}
for additional context.

If, instead, the \citet{Fafchamp_Gubert_JDE07} dyadic robust variance-covariance
estimator is used to construct standard errors (see (\ref{eq: FG_Variance_Estimate})
earlier), I get
\[
\mathbb{\hat{E}}\left[\left.Y_{ij}\right|W_{i},W_{j}\right]=\exp\left(\begin{array}{c}
-5.688\\
\left(3.6781\right)
\end{array}+\begin{array}{c}
0.9047\\
\left(0.1319\right)
\end{array}\begin{array}{c}
\mathcal{\mathtt{lyex}}\\
\\
\end{array}+\begin{array}{c}
0.8941\\
\left(0.1345\right)
\end{array}\begin{array}{c}
\mathtt{lyim}\\
\\
\end{array}+\begin{array}{c}
-0.5676\\
\left(0.2191\right)
\end{array}\begin{array}{c}
\mathtt{ldist}\\
\\
\end{array}\right).
\]
Standard errors which account for dependence across dyads sharing
an agent in common are approximately twice those which ignore such
dependence.

\subsection*{Monte Carlo experiment}

Next I report on a small Monte Carlo experiment to illustrate the
properties of inference methods based on the different variance-covariance
estimates described above. I set $N=200$ and generate outcome data
for all $N\left(N-1\right)$ ordered pairs of agents according to
the outcome model:
\[
Y_{ij}=\exp\left(\theta_{1}R_{ij}+\theta_{2}W_{2i}+\theta_{2}W_{2j}\right)A_{i}A_{j}U_{ij}
\]
Here $A_{i}$, for $i=1,...,N$, is a sequence of i.i.d. log normal
random variables, each with mean 1 and scale parameter $\sigma_{A}$;
$U_{ij}$ for $i=1,...,n$ with $n=N(N-1)$ is also sequence of i.i.d.
log normal random variables, each with mean 1 and scale parameter
$\sigma$. 

Each agent is uniformly at random assigned a location on the unit
square, ($W_{1i},W_{2i}$), $R_{ij}=\sqrt{\left(W_{1i}-W_{1j}\right)^{2}+\left(W_{2i}-W_{2j}\right)^{2}}$
equals the distance between agents $i$ and $j$ on that square; $W_{3i}$
is a standard uniform random variable. I set $\theta_{1}=-1$, $\theta_{1}=-1/2$
and $\theta_{3}=1/2$. I set $\sigma=1$ and $\sigma_{A}=1/4$. This
generates moderate, but meaningful, dependence across any two dyads
sharing at least one agent in common.

\begin{table}
\caption{\label{tab: Coverage}Coverage of different confidence intervals with
dyadic data}

\begin{centering}
\begin{tabular}{|c|c|c|}
\hline 
 & i.i.d. & dyadic clustered\tabularnewline
\hline 
\hline 
$\theta_{1}$ & 0.789 & 0.950\tabularnewline
\hline 
$\theta_{2}$ & 0.520 & 0.942\tabularnewline
\hline 
$\theta_{3}$ & 0.556 & 0.941\tabularnewline
\hline 
\end{tabular}
\par\end{centering}
\uline{Notes:} Actual coverage of nominal 0.95 confidence intervals.
The data generating process is as described in the text. Coverage
estimates are based upon 1,000 simulations. Intervals are Wald-type;
constructed by taking the coefficient point estimate and adding and
subtracting 1.96 times a standard error estimate. For the the ``i.i.d.''
column this standard error is based upon the assumption of independence
across dyads (see equation (\ref{eq: iid_standard_errors})). In the
``dyadic clustered'' column standard errors which account for dependence
across pairs of dyads sharing an agent in common are used (see equation
(\ref{eq: FG_Variance_Estimate})).
\end{table}

Table \ref{tab: Coverage} reports Monte Carlo estimates of confidence
interval coverage (the nominal coverage of the intervals should be
0.95). These estimates are based upon 1,000 simulated datasets. The
coverage properties of two intervals are evaluated. The first is a
Wald-based interval which uses standard errors constructed from (\ref{eq: iid_standard_errors}).
This corresponds to assuming independence across dyads or ``clustering
on dyads''. Confidence intervals constructed in this way are routinely
reported in, for example, the trade literature. The coverage of these
intervals is presented in first column of Table \ref{tab: Coverage}.
The second interval is based on the Fafchamps-Gubert variance estimate
(see (\ref{eq: FG_Variance_Estimate}) above). The coverage of these
intervals, which do take into account dependence across pairs of dyads
sharing an agent in common, are reported in column two of the table.

In the experiment, the intervals which do not appropriately account
fo dyadic clustering, drastically undercover the truth, whereas those
based on the variance estimator outline above have actual coverage
very close to 0.95. While there is no doubt additional work to be
done on variance estimation and inference in the dyadic context, a
preliminary suggestion is to report standard errors and confidence
intervals based upon equation (\ref{eq: FG_Variance_Estimate}) of
the previous section. These intervals perform well in the simulation
experiment, while those which ignore dyadic dependence, are not recommended.

\section{\label{sec: Further-reading}Further reading}

Although the use of gravity models by economists dates back to \citet{Tinbergen_SWE62},
discussions of how to account for cross dyad dependence when conducting
inference have been rare. \citet[Chapter 7]{Kolaczyk_NetBook09},
in his widely cited monograph on network statistics, discusses logistic
regression with dyadic data. He notes that standard inference procedures
are inappropriate due to the presence of dyadic dependence, but is
unable to offer a solution due to the lack of formal results in the
literature (available at that time).

\citet{Fafchamp_Gubert_JDE07} proposed a variance-covariance estimator
which allows for dyadic-dependence. Their estimator coincides with
the bias-corrected one discussed in \citet{Graham_HBE18} and is the
one recommended here. Additional versions (and analyses) of this estimator
are provided by \citet{Cameron_Miller_WP14} and \citet{Aronow_et_al_PA17}.
A special case of the \citet{Fafchamp_Gubert_JDE07} variance estimator
actually appears in \citet{Holland_Leinhardt_SM76} in the context
of an analysis of subgraph estimation. \citet{Snijders_Borgatti_C99}
suggested using the Jackknife for variance estimation of network statistics.
Results in, for example, \citet{Callaert_Veeraverbeke_AS81} and the
references therein, suggest that this estimate is (almost) numerically
equivalent to $\hat{\Sigma}_{1}$ defined above.

Aldous' \citeyearpar{Aldous_JMA81} representation result evidently
inspired some work on LLNs and CLTs for so called dissociated random
variables and exchangeable random arrays \citep[e.g.,][]{Eagleson_Weber_MP78}.
The influence of this work on empirical practice appears to have been
minimal. \citet{Bickel_et_al_AS11}, evidently inspired by the variance
calculations of \citet{Picard_et_al_JCB08}, but perhaps more accurately
picking up where \citet{Holland_Leinhardt_SM76} stopped (albeit inadvertently),
present asymptotic normality results for subgraph counts. Network
density, which corresponds to the mean $\left[N\left(N-1\right)\right]^{-1}\sum_{i\neq j}Y_{ij}$
when $Y_{ij}$ is binary, is the simplest example they consider and
also prototypical for understanding regression. The limit theory sketched
hear was novel at the time of drafting, but substantially related
results -- independently derived -- appear in \citet{Menzel_arXiv17}
and \citet{Davezies_et_al_arXiv2019}. Both of these papers also present
bootstrap procedures appropriate for network data. The \citet{Menzel_arXiv17}
paper focuses on the important problem of graphon degeneracy. This
occurs when the graphon only weakly varies in $U_{i}$ and $U_{j}$;
degeneracy effects rates of convergence and limit distributions. \citet{Graham_Niu_Powell_WP2019}
present results on kernel density estimation with dyadic data. \citet{Tabord-Meehan_JBES18}
showed asymptotic normality of dyadic linear regression coefficients
using a rather different approach.

\appendix

\section{\label{app: Derivations}Derivations}

Expression (\ref{eq: SIGMA1}) of the main text is an implication
of calculations like
\begin{align*}
\mathbb{V}\left(\bar{s}_{1}^{e}\left(W_{1},U_{1}\right)\right) & =\mathbb{E}\left[\mathbb{E}\left[\left.\bar{s}\left(W_{1},U_{1},W_{2},U_{2}\right)\right|W_{1},U_{1}\right]\mathbb{E}\left[\left.\bar{s}\left(W_{1},U_{1},W_{2},U_{2}\right)\right|W_{1},U_{1}\right]'\right]\\
 & =\mathbb{E}\left[\mathbb{E}\left[\left.\bar{s}\left(W_{1},U_{1},W_{2},U_{2}\right)\right|W_{1},U_{1}\right]\mathbb{E}\left[\left.\bar{s}\left(W_{1},U_{1},W_{3},U_{3}\right)\right|W_{1},U_{1}\right]'\right]\\
 & =\mathbb{E}\left[\mathbb{E}\left[\left.\bar{s}\left(W_{1},U_{1},W_{2},U_{2}\right)\bar{s}\left(W_{1},U_{1},W_{3},U_{3}\right)'\right|W_{1},U_{1}\right]\right]\\
 & =\mathbb{E}\left[\bar{s}\left(W_{1},U_{1},W_{2},U_{2}\right)\bar{s}\left(W_{1},U_{1},W_{3},U_{3}\right)'\right]\\
 & =\Omega_{12,13}.
\end{align*}
The second equality immediately above follows because $\left.W_{2},U_{2}\right|W_{1},U_{1}\overset{D}{=}\left.W_{3},U_{3}\right|W_{1},U_{1}\overset{D}{=}W_{2},U_{2}$,
the third by independence of $\bar{s}\left(W_{1},U_{1},W_{2},U_{2}\right)$
and $\bar{s}\left(W_{1},U_{1},W_{3},U_{3}\right)$ \emph{conditional}
on $W_{1},U_{1}$, and the fourth by iterated expectations.

\bibliographystyle{apalike2}
\bibliography{../Reference_BibTex/Networks_References}

\end{document}